\documentclass[12pt,epsf]{article}
\usepackage{graphicx}
\usepackage{floatflt}
\usepackage{amsmath}
\usepackage{amssymb}

\addtocounter{page}{0} \tolerance=5000

\newcommand{\beq}{\begin{equation}}
\newcommand{\eeq}{\end{equation}}
\newcommand{\beqn}{\begin{eqnarray}}
\newcommand{\eeqn}{\end{eqnarray}}

\newcommand{\BF}[1]{\mbox{\mathversion{bold}$#1$}}

\begin{document}
\begin{center}
{\Large \bf Can $CP$-violation be observed \\
\vspace{3mm} in heavy-ion collisions?}
\end{center}

\vspace{3mm}

\begin{center}
I.B.\,Khriplovich\footnote{khriplovich@inp.nsk.su} and
A.S.\,Rudenko\footnote{a.s.rudenko@inp.nsk.su}
\\Budker
Institute of Nuclear Physics\\ 630090 Novosibirsk, Russia
\end{center}

\vspace{3mm}

\begin{abstract}
We demonstrate that, at least at present, there is no convincing way to detect
$CP$-violation in heavy-ion collisions.
\end{abstract}

\vspace{3mm}

A few years ago the idea was put forward according to which in the hot and dense
matter created in the collisions of ultrarelativistic heavy nuclei, metastable
regions may form with nonvanishing values of the $\theta$-term $\theta \widetilde{F}
F$ locally violating $CP$-invariance \cite{kpt}. Certainly, the experimental
observation of this local $CP$-violation would be extremely interesting. Recently,
an extended workshop \cite{con} was dedicated to the problem.

To search for the effect, a few $P$- and $T$-odd correlations of momenta of the
pions produced in the collisions have been proposed, for instance, \cite{kp}:
\begin{equation} \label{cor}
\left(\sum_{\pi^+,\,\pi^-} \frac{\mathbf{p}_+}{|\mathbf{p}_+|} \times
\frac{\mathbf{p}_-}{|\mathbf{p}_-|}\right)\left(\sum_{\pi^+} \mathbf{p}_+ -
\sum_{\pi^-} \mathbf{p}_-\right);
\end{equation}
here $\mathbf{p}_{\pm}$ are the momenta of $\pi^{\pm}$-mesons, respectively.

However, numerical simulations \cite{fi} have demonstrated that to observe momenta
correlations on the level predicted by the model of local $CP$-violation, too high
statistics is required. We will come back to such correlations below.

\begin{floatingfigure}[l]{60mm}
\includegraphics[scale=1.3]{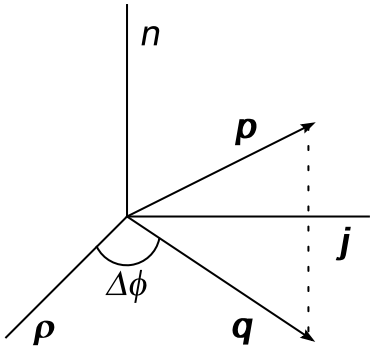}
\caption {Projections of particle momentum onto coordinate axes (notations are
defined in the text)} \label{fig:1} \vspace{2mm}
\end{floatingfigure}

One more correlation, $\langle\cos(\Delta\phi_a + \Delta\phi_b)\rangle$, was
proposed and considered in \cite{vo}. Here, the cosine is averaged over all charged
particles ($a$, $b$ denote electric charge $+$ or $-$) in each event, as well as
over all events themselves. In this expression $\Delta\phi_a$, $\Delta\phi_b$ are
the azimuthal angles of the two-dimensional projections $\mathbf{q}_a$,
$\mathbf{q}_b$ of momenta $\mathbf{p}_a$, $\mathbf{p}_b$ of particles $a$ and $b$,
respectively, onto the plane formed by the impact parameter \BF{\rho} and the
angular momentum $\mathbf{j}\,$; this plane is orthogonal to the collision axis $n$
(see the figure). This correlator is widely discussed, sometimes however with the
claim that its nonvanishing value is evidence of $P$- and $CP$-violation.

This correlation has been investigated experimentally \cite{st}. In the case when
particles $a$ and $b$ have opposite charges, the experimental results for the
correlation $\langle\cos(\Delta\phi_a + \Delta\phi_b)\rangle$ are reasonably well
reproduced by model simulations. However, there is no such agreement at all when
particles $a$ and $b$ have same charges. In fact, the analogous disagreement takes
place for some other, quite common, correlations. Most probably, the reason of the
disagreements is some shortcomings of the model simulations themselves.

But let us come back to the correlator $\langle\cos(\Delta\phi_a +
\Delta\phi_b)\rangle$. It can be conveniently rewritten as follows:
\begin{equation*}
\langle(\cos\Delta\phi_a \cos\Delta\phi_b+\sin\Delta\phi_a \sin\Delta\phi_b) -
2\sin\Delta\phi_a \sin\Delta\phi_b\rangle\,.
\end{equation*}
In this expression, the term in brackets is proportional to the scalar product of
the two-dimensional projections $\mathbf{q}_{a,\,b}$ of the particle momenta onto
the plane $(\rho j)$, and the last term is proportional to the product of the
projections of the particle momenta onto the $j$ axis. Thus, this correlator can be
conveniently presented as
\begin{equation} \label{corr}
\left<\,\left[(\mathbf{p}_{a}\mathbf{p}_{b}) -
(\mathbf{p}_{a}\mathbf{v})(\mathbf{p}_{b}\mathbf{v})\right] -
2\,(\mathbf{j}\,\mathbf{p}_{a})(\mathbf{j}\,\mathbf{p}_{b})\,\right>.
\end{equation}
Here, $\mathbf{j}$ and $\mathbf{v}$ are unit vectors directed along the total
angular momentum of the system and along the velocity of one of the beams,
respectively; $\mathbf{p}_{a,\,b}$ are the particle momenta.

Obviously, the discussed correlator is both $P$- and $T$-even, and therefore it has
by itself no direct relation to the problem of possible $P$- and $CP$-violation in
heavy-ion collisions.

On the other hand, correlator (\ref{corr}), being dependent on $(j_m j_n + j_n
j_m)$, allows one to find the axis along which the total angular momentum in a given
event is oriented.

However, nobody knows (at least, at present) how to measure the direction of the
vector $\mathbf{j}$\,, as well as that of the impact parameter $\BF{\rho}$. To
demonstrate it, let us consider the particle distribution in the azimuthal angle
$\Delta\phi$ (see the figure) \cite{vz}:
\begin{eqnarray}
\frac{dN}{d\phi} \sim 1 &+& 2 v_1 \cos(\Delta\phi) + 2 v_2
\cos(2\Delta\phi) + ... +\\
&+&  2 a_1 \sin(\Delta\phi) + 2 a_2 \sin(2\Delta\phi) + ...\nonumber
\end{eqnarray}
For our purpose it can be conveniently rewritten as
\begin{equation}
\frac{dN}{d\phi} \sim 1 + 2 v_1 (\BF{\rho}\,\mathbf{p})/(\rho q) + ...
 +  2 a_1 (\mathbf{j}\,\mathbf{p})/q + ...
\end{equation}
In fact, the $P$-odd correlator $a_1(\mathbf{j}\,\mathbf{p})$ (and those of higher
odd orders in $(\mathbf{j}\,\mathbf{p})$) is in principle measurable in the
discussed experiments \cite{st}. Still, with certainly measurable particle momentum
$\mathbf{p}$, one can fix in this way the direction of the product $a_1\mathbf{j}$
only, but not the direction of $\mathbf{j}\,$ itself: to this end, one should know
the sign of $a_1$. The same situation takes place with the vector $\BF{\rho}$. The
$P$-even correlator $v_1(\BF{\rho}\,\mathbf{p})$ is also measurable. But here as
well one can fix the direction of the product $v_1 \BF{\rho}$ only, but not the
direction of $\BF{\rho}\,$ itself.

And finally, the analogous line of reasoning applies to the idea \cite{star,du} of
measuring the global polarization of $\Lambda$ hyperons created in the heavy-ion
collisions. It consists in looking for correlation of the $\Lambda$ hyperons
polarization $\BF{\zeta}$ with the direction of the system angular momentum
$\mathbf{j}\,$: the sign of the corresponding ``coupling constant'' $\alpha$ in the
correlation $\alpha(\BF{\zeta}\mathbf{j})\,$ cannot be found independently.

Coming back to the $P$- and $T$-odd correlations, even their detection on the level
$\lesssim 10^{-3}$ in the heavy-ion collisions would not mean by itself that
$CP$-violation takes place. Indeed, at the discussed energies $\sim 100$
GeV/nucleon, the effects of parity violation, due to the exchange by the $W$- and
$Z$-bosons, can be estimated as $\alpha/\pi \sim 10^{-3}$. Then, the rescattering of
produced hadrons due to the strong interactions among them, in particular the final
state interaction, transforms these $P$-odd correlations into $P$- and $T$-odd ones
(see in this connection \cite{fi,zh,bfz}), roughly on the same level of $10^{-3}$.

On the other hand, irrespective of the idea of the spontaneous $CP$-violation, one
should expect that the $P$-odd correlation $a_1(\mathbf{j}\,\mathbf{p})$ will be
regularly present, roughly on the mentioned level of $10^{-3}$.

At last, as one more signature of possible $CP$-violation in heavy ion collisions,
the $CP$-forbidden decays $\eta, \eta^\prime \rightarrow \pi\pi$ were discussed in
\cite{kpt,zh, bfz,sh}. However, it is far from being clear whether one can reconcile
the demands arising here. On the one hand, the discussed decays should occur in a
sufficiently hot and dense medium where nonvanishing values of the $\theta$-term
$\theta \widetilde{F} F$, locally violating $CP$-invariance, do exist outside and/or
inside $\eta$\,-, $\eta^\prime$-mesons. On the other hand, the medium should be
sufficiently rarefied, so that one could talk sensibly about distinct mesons such as
$\eta, \eta^\prime$, and $\pi$. In addition, the peaks in the invariant mass of the
produced pions should be strongly smoothed out due to the rescattering in the
hadronic medium. We note also that there are strong disagreements among quantitative
predictions for the magnitude of this effect made in different models: the estimates
for the fraction of $\eta$'s decaying via forbidden channels vary from $\sim
10^{-1}$ \cite{zh} to $\sim 10^{-3}$~\cite{sh}.

\begin{center} *** \end{center}
We are grateful to A.E.\,Bondar, D.E.\,Kharzeev, V.A.\,Novikov, L.B.\,Okun,
E.V.\,Shuryak, S.A.\,Voloshin, M.I.\,Vysotsky, and W.A.\,Zajc for their interest to
our work and useful discussions. The work was supported in part by the Russian
Foundation for Basic Research through Grant no. 08-02-00960-a and by Dmitry Zimin's
Dynasty Foundation.

\end{document}